\documentclass[10pt,conference]{IEEEtran}
\usepackage{xspace}
\xspaceaddexceptions{\%}
\xspaceremoveexception{-}

\usepackage[braket,qm]{qcircuit}

\usepackage{xcolor}

\usepackage{graphicx}
\usepackage{subcaption}
\usepackage{booktabs}
\usepackage{multirow}
\usepackage{tabularx}
\usepackage[hang,flushmargin]{footmisc} % Align left footnotes

\usepackage{amsmath,amssymb,amsfonts,amsthm}

\usepackage{enumitem}
\setlist[enumerate]{leftmargin=*}
\setlist[itemize]{leftmargin=*}

\usepackage{url}
\usepackage{algorithm}
\usepackage[colorlinks]{hyperref} % Enable it for the IEEE template
\usepackage[capitalise,noabbrev]{cleveref} % Must be loaded *after* hyperref
\usepackage{algorithmic}

\usepackage{textcomp}

\def\BibTeX{{\rm B\kern-.05em{\sc i\kern-.025em b}\kern-.08em
    T\kern-.1667em\lower.7ex\hbox{E}\kern-.125emX}}

\PassOptionsToPackage{%
  backend=bibtex,bibencoding=utf8,
  language=auto, %
  style=numeric, % numeric=[1,2,3,4] vs numeric-comp=[1-4]
  sorting=none, % entries are processed in citation order
  giveninits=true, % abbreviate given/first names
  maxnames=1, % maximum number of authors before et al.
  maxbibnames=10, % default: 10, et al.
  backref=false,%
  natbib=true, % natbib compatibility mode (\citep and \citet still work)
  doi=true,
  url=true
}{biblatex}
\usepackage{biblatex}
\usepackage{balance}

\usepackage{tikz}
\newcommand{\circnum}[1]{%
\tikz[baseline=(char.base)]{
\node[shape=circle,draw,inner sep=1pt] (char) {#1};}}

\begin{document}

\title{Quantum Mutant Equivalence via Transpilation}

\author{
\IEEEauthorblockN{José Campos}
\IEEEauthorblockA{
%\textit{dept. name of organization (of Aff.)} \\
\textit{Faculdade de Engenharia, Universidade do Porto} \\
\textit{LASIGE, Faculdade de Ciências, Universidade de Lisboa}\\
Porto, Portugal \\
jcmc@fe.up.pt}
\and
\IEEEauthorblockN{Andriy Miranskyy}
\IEEEauthorblockA{
%\textit{Department of Computer Science} \\
\textit{Toronto Metropolitan University}\\
Toronto, Canada \\
avm@torontomu.ca}
}

\maketitle

% ------------------------------------------------------------------- Abstract

\begin{abstract}
Mutation testing evaluates test suite quality by introducing artificial faults (mutants) and checking whether tests detect (kill) them.  A central challenge is the equivalent mutant problem: some mutants are syntactically different from the original program but semantically identical to it and therefore cannot be killed by any test.  If left unidentified, such mutants waste testing effort and distort mutation scores.
In quantum software, mutation testing is increasingly used, but the equivalent mutant problem remains unsolved.  A recent study generated more than 700,000 quantum circuit mutants and found that roughly half survived the available tests, making it unclear whether these survivors reflect weak tests or semantic equivalence.
We propose Transpiler-Based Equivalence~(TBE), 
a lightweight approach that identifies equivalent quantum mutants by transpiling original and mutated circuits under the same configuration and comparing their resulting OpenQASM code.  We evaluate TBE on 348,299 surviving mutants, 92,011 of which are equivalent; TBE identifies 29,536 of them (32.1\%) as equivalent while achieving 100\% precision and 82\% accuracy.

\end{abstract}


% ------------------------------------------------------------------- Introduction

\section{Introduction}\label{sec:intro}

Mutation testing evaluates test suite quality by introducing small syntactic changes, or mutants, into a program and checking whether existing tests detect them~\cite{10472898,QMutPy,Muskit}.  The resulting mutation score is commonly used as a proxy for test thoroughness.  However, mutation testing faces the equivalent mutant problem: some mutants are syntactically different from the original program but semantically identical to it.  Such mutants cannot be killed by any test, yet they may be incorrectly counted as surviving mutants, thereby distorting mutation scores and wasting testing effort.  Because exact equivalence detection is undecidable in general~\cite{10.1007/BF00625279}, practical techniques usually aim to identify a useful subset of equivalent mutants rather than solve the problem completely~\cite{Papadakis2015}.

This problem also arises in quantum software, where programs are often represented as circuits that transform input quantum states through sequences of gates.  Two quantum circuits may use different gate sequences while implementing the same overall transformation, and a mutation may therefore preserve circuit semantics even when it changes the gate-level representation.  For example, inserting a gate followed by its inverse yields an equivalent circuit.

A recent large-scale study~\cite{UsandizagaMutation2025} generated 723,079 mutants from real-world quantum circuits and found that 348,299 (48.2\%) survived the available tests.  These surviving mutants may indicate either inadequate tests or semantic equivalence.  The authors noted that ``there is no sufficient understanding of quantum mutations'' and emphasized the need to analyze which mutants are difficult to detect.  This paper addresses that gap by automatically identifying equivalent mutants among the surviving mutants.

To address this problem, we propose Transpiler-Based Equivalence~(TBE), a lightweight approach for identifying equivalent quantum mutants.  TBE transpiles the original and mutated circuits (i.e., converts both circuits into hardware-native gates and optimizes them), under the same configuration and compares their resulting OpenQASM code.  When the outputs match, TBE conservatively classifies the mutant as equivalent; otherwise, the mutant remains for standard mutation analysis.

\smallskip
\noindent
The contributions of this paper are as follows:
\begin{enumerate}[leftmargin=*]
  \item[\small{$\bigstar$}] A novel approach, Transpiler-Based Equivalence~(TBE), that
  detects equivalent mutants in quantum programs.

  \item[\small{$\bigstar$}] A preliminary study evaluating TBE's effectiveness on 348,299 surviving mutants previously generated~\cite{UsandizagaMutation2025} for the well-known MQT~benchmark~\cite{mqt}.
\end{enumerate}

% ------------------------------------------------------------------- TBE

\section{Transpiler-Based Equivalence (TBE)}\label{sec:tte}

Transpiler-Based Equivalence~(TBE) is a lightweight approach for identifying equivalent quantum mutants.\footnote{TBE is built on top of Qiskit framework and therefore uses Qiskit's transpiler procedure.  Nevertheless, the TBE approach is framework agnostic: other frameworks, such as Cirq, PennyLane, Braket SDK, TKET, and PyQuil, provides similar procedures like optimization or compilations procedures.}

Let $C_O$ denote the original circuit and $C_M$ denote a mutant.  We write $\tau_{\theta}(C)$ for the result of transpiling circuit $C$ under a transpiler configuration $\theta$, and $\mathrm{QASM}(C)$ for the textual OpenQASM representation of circuit $C$.  TBE transpiles both circuits, $C_O$ and $C_M$, using the same configuration and compares the resulting OpenQASM representations as:
\begin{equation} \label{eq:qasm_comparison}
  \mathrm{QASM}[\tau_{\theta}(C_O)] =
  \mathrm{QASM}[\tau_{\theta}(C_M)].
\end{equation}
If the two representations are identical, TBE classifies the mutant as equivalent and removes it from the mutation-analysis workload.
This can occur when a mutation is optimized away, merged with neighboring gates, or reduced to the same basis gate sequence as the original circuit.  The approach is analogous to compiler-based equivalent-mutant detection in classical mutation testing~\cite{Papadakis2015}, but it uses quantum transpilation.

TBE is conservative: equality of the transpiled representations is sufficient evidence for equivalence under the chosen transpiler configuration, whereas inequality only means that equivalence was not detected.
The approach assumes that the transpiler preserves circuit semantics, making TBE appropriate when the transpiler is trusted as an optimizer, but not when the transpiler itself is under test.  Because TBE relies on standard transpilation rather than explicit state-vector or matrix representations, its practical scalability is tied to the scalability of the underlying transpiler.

% ------------------------------------------------------------------- Study

\section{Empirical Study}\label{sec:study}

In this section, we evaluate the effectiveness of TBE at identifying equivalent quantum mutants. In particular, we investigate the following research question (\textbf{RQ}): \textit{How effective is TBE at identifying surviving mutants as equivalent?}
To answer this question, we compare TBE classifications against a state-vector baseline.  Specifically, we assess how closely the equivalent/unknown classifications produced by TBE match those obtained through the direct state-vector comparison described below.

\subsection{Baseline}\label{sec:study:baseline}

We use state-vector simulation as the reference classification for the RQ.  For each original--mutant pair, we simulate\footnote{This can be done using \texttt{Statevector.from\_instruction(circuit)} in the Qiskit framework~\cite{javadi2024quantum}.} the output state of the original circuit, $\ket{\psi_O}$, and the output state of the mutated circuit, $\ket{\psi_M}$, using the same input state.  The mutant is classified as equivalent when the two output states are equal within tolerance, up to the global phase.

State-vector simulation provides a precise baseline~\cite{quantumCentricTest} for the relatively small benchmark circuits studied here, but it scales exponentially with the number of qubits $n$ because the state vector has size $2^n$.  We propose and evaluate TBE as a cheaper alternative that can identify some equivalent mutants without explicit simulation.

\subsection{Subjects}\label{sec:study:subjects}

To evaluate TBE, we use the 348,299 surviving mutants previously generated by \citet{UsandizagaMutation2025} for 375 quantum circuits (from the MQT~benchmark~\cite{mqt}), ranging from 2 to 30 qubits.  These mutants survived the original mutation analysis and therefore represent candidates for semantic equivalence.

\subsection{Experimental Methodology}\label{sec:study:procedures}

\circnum{1} For each mutant, we compute the state vector of the original and mutant circuits and compare them using Qiskit's \texttt{equiv} function with its default tolerance of $\varepsilon = 1 \times 10^{-8}$.  Mutants whose state vectors match within the tolerance are classified as equivalent.  This state-vector comparison serves as the reference classification for semantic equivalence, against which we evaluate TBE's classifications.  

\circnum{2} We apply TBE to the same original--mutant pairs.  As defined in \Cref{eq:qasm_comparison}, TBE transpiles both circuits using the same transpiler configuration $\theta$ and compares the resulting OpenQASM representations.  In this study, we use the IBM Qiskit transpiler without device-specific noise models, in order to isolate semantic equivalence at the circuit level from hardware-dependent execution effects.  We configure the transpiler with:
\begin{itemize}
  \item Five different sets of basis gates~\cite{StefanoSmellsStudy}, denoted by $G$:
  \begin{enumerate}
    \item IBM Perth (\texttt{cx}, \texttt{id}, \texttt{rz}, \texttt{sx}, and \texttt{x})

    \item IBM Sherbrooke (\texttt{ecr}, \texttt{id}, \texttt{rz}, \texttt{sx}, and \texttt{x})

    \item QSmell~\cite{ChenSmells} (\texttt{u1}, \texttt{u2}, \texttt{u3}, \texttt{rz}, \texttt{sx}, \texttt{x}, \texttt{cx}, and \texttt{id})

    \item RPCX (\texttt{rx}, \texttt{ry}, \texttt{rz}, \texttt{p}, and \texttt{cx})

    \item Simple (\texttt{cx} and \texttt{u3})
  \end{enumerate}

  \item Optimization level~3, the maximum optimization level supported by the Qiskit transpiler, denoted by $l$.
\end{itemize}

Because optimization level~3 is nondeterministic, we repeat TBE $R=100$ times.  We treat each repetition as one complete execution of the TBE procedure under a fixed transpiler seed.  Let $s_r$ denote the transpiler seed that controls randomized transpiler choices used in repetition $r$, for $r \in \{1,\ldots,R\}$.  We therefore write the transpilation step as $\tau_{\theta}(C)$, where $\theta$, which denotes the transpiler configuration, is initialized as the triple $(G, l, s_r)$.
For each repetition $r$, TBE classifies every original--mutant pair by checking whether
$
  \mathrm{QASM}[\tau_{(G, l, s_r)}(C_O)] =
  \mathrm{QASM}[\tau_{(G, l, s_r)}(C_M)].
$
For a given original--mutant pair and repetition, both circuits are transpiled using the same configuration so that differences in their resulting OpenQASM representations are not caused merely by different randomized transpiler choices.  If the resulting transpiled OpenQASM representations are identical, TBE classifies the mutant as equivalent.
All other transpiler settings use the defaults in Qiskit~v2.4.1.

\circnum{3} We compare the TBE classifications against the state-vector baseline. For each repetition $r$, we compute the number of true positives (TP), false positives (FP), true negatives (TN), and false negatives (FN), where the positive class corresponds to equivalent mutants. From these counts, we report accuracy (acc.), precision (prec.), recall (rec.), F1-score, and Matthews Correlation Coefficient~(MCC)\footnote{MCC is  useful for imbalanced classes, where accuracy and F1-score can be misleading~\cite{yao2021imapct}.}.
Across the 100 repetitions, we report median metric values,
confidence intervals using bootstrapping at 95\% confidence level, and standard deviation.
These intervals quantify sensitivity to randomized transpilation rather than sampling uncertainty over mutants.

\smallskip
\textit{Runtime environment:}
The experiments were carried out on
the Deucalion supercomputer.
Each experiment, i.e., the execution of an approach, either state-vector or TBE, on the original and mutant circuit, was allocated 94~GB of memory and 1~CPU core.  The total computational cost for running all experiments was approximately 112~CPU-years.

\subsection{Threats to Validity}\label{sec:study:threats}

\subsubsection{Internal validity}
The main internal threat concerns the correctness of the implementation and configuration of the equivalence checks.  Errors in parsing OpenQASM programs, invoking the transpiler, comparing generated QASM code, or computing state vectors could affect mutant classification.  To mitigate this threat, we use the same pipeline across all mutants and compare TBE against a state-vector baseline configured with a default numerical tolerance $\varepsilon$.  

A second internal threat is transpiler nondeterminism, especially at optimization level~3.  We mitigate this threat by repeating the TBE experiments 100 times with fixed transpiler seeds while holding all other transpiler settings constant.  The resulting metric distributions quantify sensitivity to randomized transpilation rather than sampling uncertainty over mutants.

\subsubsection{External validity}
Our results may not generalize to all quantum programs, benchmarks, or mutation operators.  The study uses the 348,299 surviving mutants previously generated by \citet{UsandizagaMutation2025} for the MQT benchmark.  Although MQT is widely used in quantum software engineering research, other benchmarks may contain circuits with different structures, gate sets, depths, or numbers ofqubits; other mutation tools may also produce different equivalence patterns.

The results also depend on the selected TBE configuration.  Different basis gate sets, optimization levels, or hardware backends may expose different sets of equivalent mutants.  Future work will evaluate additional configurations, including different optimization levels and  unrolling choices.

\subsubsection{Construct validity}
Construct validity concerns whether the measurements capture the intended concepts.  We use state-vector comparison as the baseline for semantic equivalence. This is appropriate for the circuits studied here, but it may not scale to larger circuits because state-vector size grows exponentially with $n$.  Because equivalence is assessed within a numerical tolerance $\varepsilon$, we treat the state-vector result as a numerical reference rather than a symbolic proof.

TBE operationalizes equivalence by comparing the OpenQASM code produced after transpiling the original and mutated circuits.  Thus, it can identify mutants whose equivalence is exposed by transpilation, but it is not a complete equivalence checker: two semantically equivalent circuits may still yield different OpenQASM outputs.  Therefore, TBE should be interpreted as a practical, conservative detector of equivalent mutants, not as a complete decision procedure.

\subsection{Results and Answer to RQ}\label{sec:study:rq1}

\begin{figure}[tb]
  \centering
  \includegraphics[width=0.95\columnwidth]{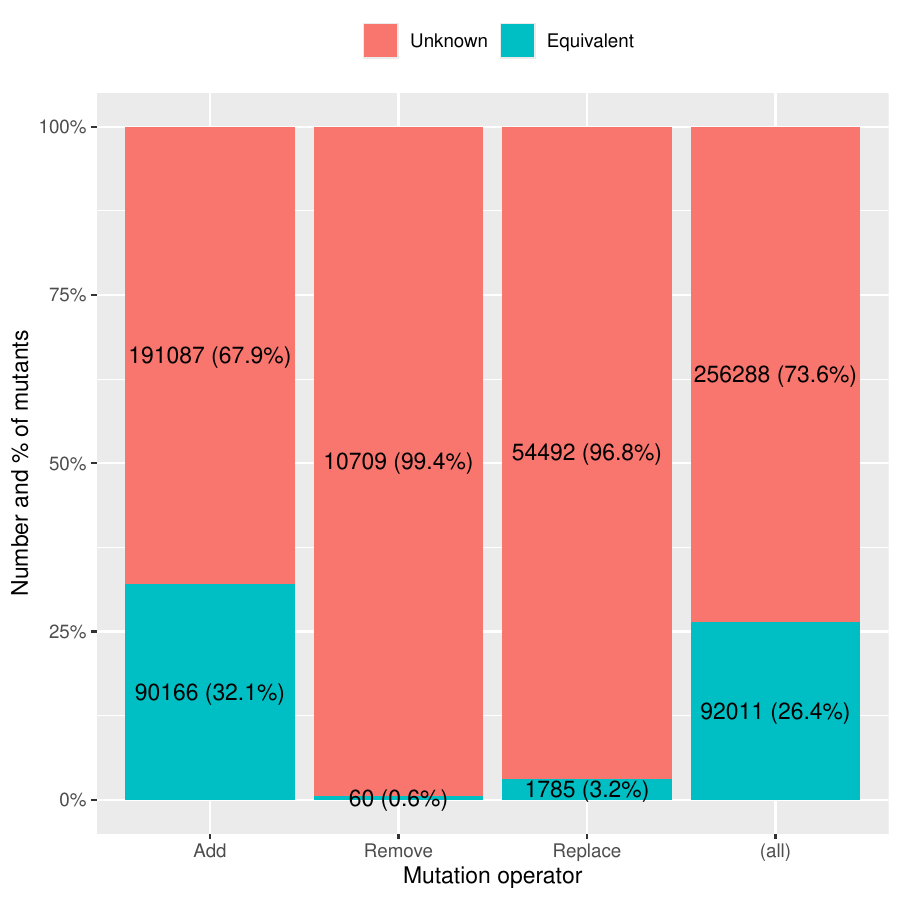}
  \caption{State-vector analysis.
  {\footnotesize The \textit{(all)} column considers all mutants generated by all mutation operators.}}
  \label{fig:survived-mutants-vs-statevector}
\end{figure}

\Cref{fig:survived-mutants-vs-statevector} reports the results of the state-vector analysis.  Overall, \textbf{92,011 (26.4\%) of the 348,299 surviving mutants} previously generated by \citet{UsandizagaMutation2025} \textbf{are equivalent}.  Most mutants classified as equivalent (90,166) are produced by mutation operators that add new gates to the original circuit.  Additionally, the state-vector results suggest that mutating a circuit by removing or replacing a gate is more likely to produce a non-equivalent mutant.

\begin{table}[tb]
\centering
\caption{TBE performance for different basis gate sets.
For all metrics, with the exception of FP and FN, higher values indicate better performance.
}\label{table:tbeperformance}
\resizebox{\columnwidth}{!}{
\begin{tabular}{@{}l|lllllllll@{}}
\toprule
Basis gates set & TP & TN & FP & FN & Acc. & Prec. & Rec. & F$_1$ & MCC \\
\midrule
IBM Perth & 29438 & 256288 & 0 & 62573 & 0.820 & 1.000 & 0.320 & 0.485 & 0.507 \\
IBM Sherbrooke & 29472 & 256288 & 0 & 62539 & 0.820 & 1.000 & 0.320 & 0.485 & 0.507 \\
QSmell & 29272 & 256288 & 0 & 62739 & 0.820 & 1.000 & 0.318 & 0.483 & 0.506 \\
RPCX & 29383 & 256288 & 0 & 62628 & 0.820 & 1.000 & 0.319 & 0.484 & 0.507 \\
Simple & 29410 & 256288 & 0 & 62601 & 0.820 & 1.000 & 0.320 & 0.484 & 0.507 \\
\bottomrule
\end{tabular}
}
\end{table}

\begin{figure}[tb]
  \centering
  \includegraphics[width=0.95\columnwidth]{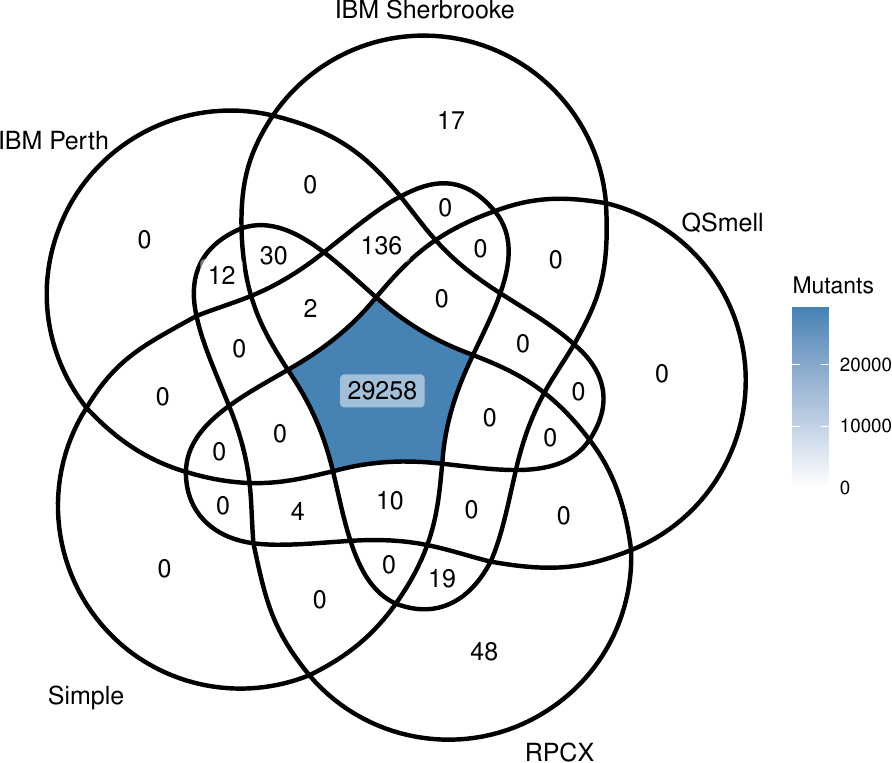}
  \caption{True positives overlap among all basis gates sets.}
  \label{fig:survived-mutants-vs-true-positive-venn}
\end{figure}

\Cref{table:tbeperformance} reports TBE's effectiveness.
First, independently of the basis gates set, \textbf{TBE performed consistently across all 100 repetitions}.  That is, it identified the same set of equivalent mutants in all repetitions, yielding zero-width confidence intervals and zero standard deviation for all metrics.

Second, \textbf{TBE produced no false positive}, i.e., every mutant classified as equivalent by TBE was also equivalent according to the state-vector baseline.

Third, \textbf{the number of true positives is similar across basis gate sets}, ranging from 29,272 with QSmell to 29,472 with IBM Sherbrooke.  
\Cref{fig:survived-mutants-vs-true-positive-venn} shows that \textbf{TBE identifies a total of 29,536 unique equivalent mutants} (32.1\% of the 92,011 state-vector-equivalent mutants).  Most of these mutants (29,258) are detected under every basis gate set, while the union of IBM Sherbrooke and RPCX detects all 29,536.

Fourth, the 256,288 true negatives (i.e., mutants classified as unknown by the state-vector baseline and TBE) suggest that 73.6\% of the surviving mutants may still be killable by some test input.
The 62,539--62,739 false negatives (i.e., mutants classified as equivalent by the state-vector baseline but classified as unknown by TBE) show TBE's main limitation: many equivalent mutants are not exposed by the selected transpilation configurations.  Future work should therefore evaluate additional configurations.

Finally, on average, \textbf{TBE achieves 0.820 accuracy, 1.000 precision, 0.320 recall, 0.484 F1-score, and 0.507 MCC}.
These results characterize TBE as a conservative detector: it \textbf{makes highly reliable positive predictions but detects only about one-third of the equivalent mutants found by the state-vector baseline}.

% ------------------------------------------------------------------- RW

\section{Related Work}\label{sec:rw}

Prior work has studied equivalent mutants in both classical and quantum mutation testing.  In classical mutation testing, compiler-based techniques identify equivalent mutants when optimization reduces the original and mutant programs to the same representation~\cite{Papadakis2015}.  TBE adapts this idea to quantum circuits by using transpilation rather than classical compiler optimization.

In quantum mutation testing, \citet{wang2022mutation} guide test generation by discounting mutants that are likely to be equivalent, whereas TBE directly filters mutants whose equivalence is exposed by transpilation. \citet{Kumar2023} compare matrix representations before hardware execution, and \citet{QGMR} use state-vector comparison to precompute equivalent gate combinations during mutant generation. TBE instead acts after mutant generation and before execution by comparing transpiled OpenQASM representations. Other recent work reduces redundant quantum machine-learning mutants~\cite{andrews2026efficient} or studies equivalence and behavioral similarity under hardware noise~\cite{fortz2026robust}; TBE focuses on circuit-level equivalence under a fixed transpilation configuration.

% ------------------------------------------------------------------- Conclusions

\section{Conclusions and Future Work}\label{sec:conclusion}

We presented TBE, a lightweight approach for detecting equivalent quantum mutants by comparing transpiled OpenQASM representations. In our empirical study, TBE detected semantically equivalent mutants with no false positives while requiring only about $\frac{1}{445}$ of the execution time and $\frac{1}{2}$ of the memory required by the state-vector baseline.
Future work will evaluate additional configurations, including different basis gate set, optimization levels, and unrolling choices,
and study duplicate mutants: mutants that are not equivalent to the original circuit but are semantically equivalent to other mutants.

% ------------------------------------------------------------------- Acks

% \section*{Acknowledgment}
% 
% TODO

%
% IEEE
%
\printbibliography

\end{document}